\documentclass[a4paper,fleqn,usenatbib]{mnras}

\usepackage{newtxtext,newtxmath}

\usepackage[T1]{fontenc}
\usepackage{ae,aecompl}

\usepackage{graphicx}	% Including figure files
\usepackage{amsmath}	% Advanced maths commands
\usepackage{amssymb}	% Extra maths symbols

\title[Subaru FOCAS IFU observations of two MaNGA lenses]
{Subaru FOCAS IFU observations of two $\boldsymbol{z}$\,$\boldsymbol{\approx}$\,0.12 strong-lensing elliptical galaxies from SDSS MaNGA\thanks{Based on data collected at Subaru Telescope, which is operated by the National Astronomical Observatory of Japan.}}
\author[Russell J. Smith et al.]{
	Russell J. Smith$^1$\thanks{E-mail: russell.smith@durham.ac.uk},
	William P. Collier$^1$,
	Shinobu Ozaki$^2$ and 
	John R. Lucey$^1$
	\\
	$^1$Centre for Extragalactic Astronomy, University of Durham, Durham DH1 3LE, United Kingdom\\
	$^2$National Astronomical Observatory of Japan, 2-21-1 Osawa, Mitaka, Tokyo 181-8588, Japan
}

\pubyear{2019}

\begin{document}
\label{firstpage}
\pagerange{\pageref{firstpage}--\pageref{lastpage}}
\maketitle

\newcommand{\ja}{J1436+4943}	
\newcommand{\jb}{J1701+3722}	
\newcommand{\oii}{[O\,{\sc ii}]}
\newcommand{\oiii}{[O\,{\sc iii}]}
\newcommand{\hb}{H$\beta$}

\begin{abstract}
We present new observations of two $z$\,=\,0.12 strong-lensing elliptical galaxies, originally discovered from the SDSS-IV MaNGA survey, using the new FOCAS IFU spectrograph on the Subaru Telescope.
For \ja, our observations confirm the identification of this system as a
multiple-image lens, in a cusp configuration, with Einstein radius $\theta_{\rm Ein}$\,=\,2.0\,arcsec.
For \jb, the improved data confirm earlier hints of a complex source plane, with different configurations evident in different emission lines. The new observations reveal a previously unseen inner counter-image to the [O\,{\sc iii}] arc found from MaNGA, leading to a smaller revised Einstein radius of $\theta_{\rm Ein}$\,=\,1.6\,arcsec.
The inferred projected masses within the Einstein apertures (3.7--4.7\,kpc) are
consistent with being dominated by stars with 
an initial mass function (IMF) similar to that of the Milky Way, and a dark matter contribution of $\sim$35\,per cent as supported from cosmological simulations. 
These results are consistent with 
`pure lensing' analyses
of lower-redshift lenses, but contrast with claims for heavier IMFs from
combined lensing-and-dynamical studies of more distant early-type galaxies.
\end{abstract}


\section{Introduction}

Strong gravitational lensing is a key tool for measuring extragalactic masses and probing the distribution of dark and luminous matter in galaxies \citep[e.g.][]{2010ARA&A..48...87T}.
 Analysing a sample of $z$\,$\sim$\,0.2 strong-lensing elliptical galaxies, 
  \cite{2010ApJ...709.1195T} found that their velocity dispersions and projected lensing masses could only be reconciled with a universal dark matter profile if the stellar mass-to-light ratio, $\Upsilon$, was larger than expected for an initial mass function (IMF) like that in the Milky Way \citep[e.g.][]{2001MNRAS.322..231K}. 
  By contrast, for the few known nearby ($z$\,=\,0.03--0.07) strong lenses, 
lensing {\it alone} constrains $\Upsilon$ to be close to the value expected from the Kroupa IMF \citep{2018MNRAS.478.1595C}.
In these low-redshift systems, the arcs are formed at smaller physical radius, minimising the contributions from dark matter, which obviates the need for dynamical information to separate dark and stellar components.
 One possible resolution of the tension between the above results, discussed by \cite{2019A&A...630A..71S}, is that steep inner gradients in $\Upsilon$ may 
inflate the central velocity dispersion, causing the combined lensing-and-dynamics analysis to assign too much mass to the stellar component and too little to the dark halo.  

Establishing well-constrained lenses at intermediate distances should help to solve
the apparent discrepancy between the nearby and distant samples.
\citet*{2015MNRAS.449.3441S} presented the first use of Integral Field Unit (IFU) spectroscopy as a tool to discover lensed line-emitting sources behind bright foreground galaxies. This method, applied to individually targeted lens candidates, has also been adapted to the VLT MUSE spectrograph  
{\citep[][and in preparation]{2015MNRAS.449.3441S,2018MNRAS.478.1595C}}.
A complementary approach is to exploit data from large multi-IFU surveys.
The SDSS-IV MaNGA survey \citep{2015ApJ...798....7B} is acquiring IFU observations for thousands of nearby galaxies, including hundreds of massive ellipticals, which are the most effective gravitational lenses.  

Two groups have searched the MaNGA data cubes for lens systems. \cite{2017MNRAS.464L..46S} (and unpublished extensions) conducted a semi-automated search in MaNGA data releases DR13--DR15, restricted to $\sim$600 galaxies with velocity dispersion $\sigma$\,$>$\,200\,km\,s$^{-1}$,  which dominate the total strong-lensing cross section. Independently, \cite{2018MNRAS.477..195T} searched all datacubes in DR13--DR14 with a fully-automated source detection method.
\cite{2017MNRAS.464L..46S} reported discovery of a lensed source behind \jb, apparently showing a near-complete Einstein ring in [O\,{\sc ii}], but only a single `arc' in  [O\,{\sc iii}]. He estimated an Einstein radius of 2.3\,arcsec, but noted that this is uncomfortably large compared to the measured velocity dispersion.  The \cite{2018MNRAS.477..195T} search yielded seven new apparent multiple-imaging systems as well as confirming identification of \jb. In a subsequent re-analysis of the data, we could not convincingly reproduce many of these systems \citep[see discussion in][]{2018MNRAS.481.2115S}, but  \ja,  which was missed in Smith's visual search, is a compelling candidate for further study.

In this {\it Letter}, we present follow-up observations of \ja\ and \jb, obtained with the new IFU mode of the FOCAS spectrograph at the 
Subaru telescope (Section~\ref{sec:obs}). The new observations have much higher spatial resolution than the MaNGA discovery data, 
and also higher spectral resolution, 
enabling a more detailed reanalysis of the two lenses
(Sections~\ref{sec:ja}--\ref{sec:jb}). Combining the measured Einstein radii with photometry and estimates for the dark matter contributions and likely stellar ages, we derive results on the total and stellar mass-to-light ratios, and constraints on the IMF through the mass-excess factor (Section \ref{sec:mlcalc}). The work is briefly summarized in Section~\ref{sec:concs}.

We adopt the  cosmological parameters  from 
\cite{2018arXiv180706209P}.

\section{Subaru FOCAS IFU observations}\label{sec:obs}

We observed \ja\ and \jb\ using the Subaru FOCAS IFU during its first night of scientific operations, on 2019 June 26.
The design and characteristics of the IFU are fully described by 
\citet[][and in preparation]{2014SPIE.9151E..49O}. Briefly, the unit is an image slicer which reformats a 10.1$\times$13.5\,arcsec$^2$ field-of-view into 23 slices of width 0.44\,arcsec, sampled with 0.2\,arcsec pixels; the IFU feeds the existing
FOCAS spectrograph \citep{2002PASJ...54..819K}.
We  observed with the VPH900 grism, providing a wavelength coverage of 7600--10600\,\AA\ over most of the field-of-view,
with spectral resolution  $R$\,$\approx$\,2800--3300, 
sampled at 0.74\,\AA\,pix$^{-1}$ 
in wavelength.

For \ja\ we acquired six dithered exposures totalling 2943\,sec integration, 
while for \jb\ a single 900\,sec exposure was obtained.
The seeing was $\sim$0.75\,arcsec FWHM for \ja, 
and $\sim$0.5\,arcsec for \jb\ (estimated from focus-sequence images of a star  taken
close to the observation time). Especially for \jb, the PSF is thus severely under-sampled in the across-slice dimension (R.A.).

The data were reduced using the observatory pipeline software. The wavelength calibration solution was obtained from the science frames themselves, using the numerous night-sky emission lines as a simultaneous reference source.

\section{SDSSJ\,143607.49+494313.2}\label{sec:ja}

\begin{figure}
\begin{center}
\vskip -1mm
\includegraphics[width=75mm]{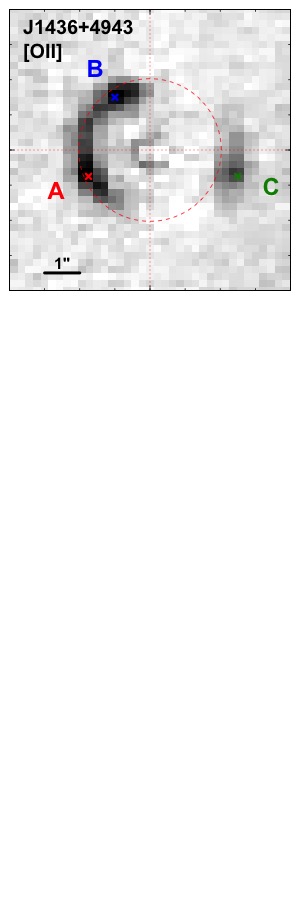}
\end{center}
\vskip -153mm
\caption{Net emission line image of the $z_{\rm s}$\,=\,1.231 \oii\ doublet in the background source of 
 MaNGA lens system \ja. 
North is at the top, east at the left, 
and axis ticks are spaced by one arcsecond. Coloured crosses mark the centres of regions used to extract the spectra in Figure~\ref{fig:j14spec}. The  dashed red circle indicates the derived Einstein radius $\theta_{\rm Ein}$\,=\,2.03\,arcsec.}
\label{fig:j14}
\end{figure}

\begin{figure}
	\begin{center}
		\vskip -3mm
		\includegraphics[width=80mm]{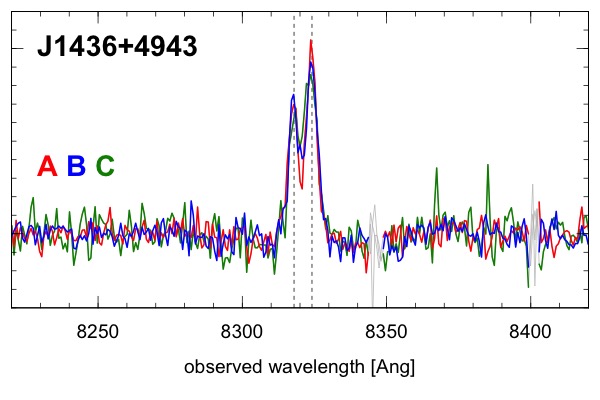}
	\end{center}
	\vskip -6mm
	\caption{Extract from the spectrum of the \ja\ system showing the $z_{\rm s}$\,=\,1.231 \oii\ doublet from the three regions identified in Figure~\ref{fig:j14}.
		Pixels with increased noise due to sky lines are shown in grey.
	}
	\label{fig:j14spec}
\end{figure}

The \ja\ system was first identified as a possible multiple-imaging system by \cite{2018MNRAS.477..195T}. The foreground galaxy is an elliptical at 
redshift $z_{\rm l}$\,=\,0.125, with velocity dispersion $\sigma$\,=\,282$\pm$10\,km\,s\,$^{-1}$
from the single-fibre SDSS spectrum \citep{2009ApJS..182..543A}. 
MaNGA revealed background \oii\ emission at $z_{\rm s}$\,=\,1.231,
 apparently forming two separate arcs east and west 
of the target galaxy \citep{2018MNRAS.477..195T}.

Figure~\ref{fig:j14} shows the net \oii\ emission line image for the source, 
from the Subaru data. The lens light has been subtracted using the neighbouring continuum, and a correction has been made to remove a radial trend in the net image (caused by gradients in spectral features in the lens galaxy).
The \oii\ image confirms the presence of an extended arc (labelled A and B) to the east,
and a western counter-image (C), suggesting a cusp lens configuration.
Figure~\ref{fig:j14spec} shows the \oii\ doublet in the 
spectra for the three labelled regions.

To derive the Einstein radius, we have fitted simple lensing models to the net \oii\ image using 
{\sc PyAutoLens} \citep{2018ascl.soft07003N,2018MNRAS.478.4738N}. The source galaxy is parameterized with an elliptical S\'ersic profile, while the lens is treated as a singular isothermal ellipsoid (SIE), with centre, ellipticity and position angle fixed to those of the lens galaxy light. Fitting for the lens normalisation, we obtain $\theta_{\rm Ein}$\,=\,2.02$\pm$0.02\,arcsec; allowing a free external shear term does not 
affect the fit. 
Fitting with a pixelised source, we find  $\theta_{\rm Ein}$\,=\,2.08$\pm$0.04 with no shear, or 2.05$\pm$0.03 with a 3$\pm$2\% shear. The reconstructed source is compact and unstructured in each case. For the calculations in Section~\ref{sec:mlcalc}, we adopt $\theta_{\rm Ein}$\,=\,2.03$\pm$0.04, averaging the results from pixelised and parametric models with shear, with the error reflecting the variation between models.

\section{SDSSJ\,170124.01+372258.1}\label{sec:jb}

\begin{figure*}
\begin{center}
\vskip -1mm
\includegraphics[width=155mm]{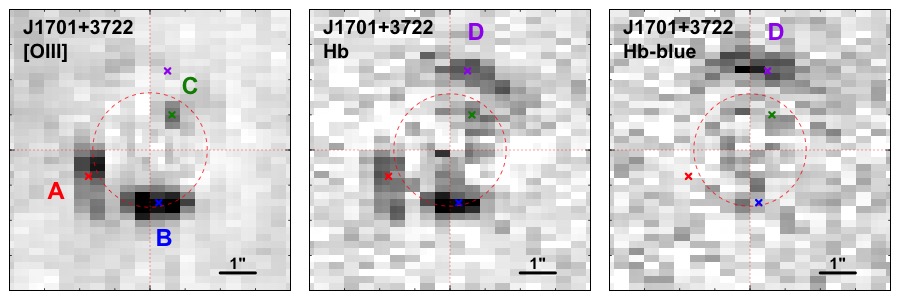}
\end{center}
\vskip -4mm
\caption{Net emission line images for the $z_{\rm s}$\,=\,0.791 background source in MaNGA lens system \jb.
	Orientation and scale are as in Figure~\ref{fig:j14}. The left and central panels are centred on \oiii\ and \hb\ for the source which forms images A, B and C. The right panel is offset to shorter wavelength, isolating the  H$\beta$-bright contaminating source, D. 
	Coloured crosses mark the centres of regions used to extract the spectra in Figure~\ref{fig:j17spec}. The dashed red  circle indicates the derived Einstein radius $\theta_{\rm Ein}$\,=\,1.63\,arcsec.}
\label{fig:j17oiii} 
\end{figure*}

\begin{figure*}
\begin{center}
\vskip -2mm
\includegraphics[width=152mm]{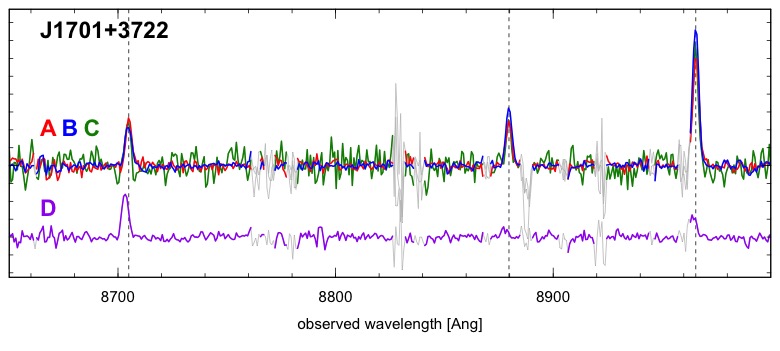}
\end{center}
\vskip -6mm
\caption{Extracts from the spectrum of the \jb\ system showing the $z_{\rm s}$\,=\,0.791
[O\,{\sc iii}] and H$\beta$ lines from the four arcs as identified in Figure~\ref{fig:j17oiii}. Regions of increased noise are shown in grey.
Note the small velocity offset ($\sim$60\,km\,s$^{-1}$) and very different line ratios in the spectrum of arc D.}
\label{fig:j17spec}
\end{figure*}

\jb\ was identified as a multiple-image lens by \cite{2017MNRAS.464L..46S},
and also recovered in the \cite{2018MNRAS.477..195T} search. 
The foreground galaxy is an elliptical at 
redshift $z_{\rm l}$\,=\,0.122, with velocity dispersion $\sigma$\,=\,258$\pm$8\,km\,s\,$^{-1}$ from the single-fibre SDSS spectrum. 
MaNGA detected \oii, \oiii, and \hb\ emission lines from a background source at 
$z_{\rm s}$\,=\,0.791.

As discussed by \cite{2017MNRAS.464L..46S}, the MaNGA data for \jb\ posed two puzzles. First, although the system exhibited a
near-complete Einstein ring in the [O\,{\sc ii}] and H$\beta$ lines, only
a single south-eastern `arc' was visible in  [O\,{\sc iii}].  Secondly, the estimated Einstein radius of $\theta_{\rm Ein}$\,=\,2.3\,arcsec
derived from the [O\,{\sc ii}] ring was substantially larger
than the $\sim$1.5\,arcsec\ predicted for 
an isothermal profile matching the stellar velocity dispersion.

Figure~\ref{fig:j17oiii} (left) shows the net emission-line image for the background [O\,{\sc iii}]  5007\,\AA\ line. 
The south-eastern arc identified from MaNGA is resolved by FOCAS into two distinct sections, labelled A and B, at $\sim$1.8\,arcsec from the lens centre.  Additionally, a probable faint counter-image, C, is now visible at smaller radius, $\sim$1.1\,arcsec. 
Turning to the \hb\ line, Figure~\ref{fig:j17oiii} (centre) shows that 
features A, B and (marginally) C are all detected, but the situation is complicated by an additional arc, D, at radius $\sim$2.4\,arcsec, which is
very faint in the \oiii\ image. 
This feature clearly corresponds to the northern arc seen in H$\beta$ and \oii\ in MaNGA. Note that the  \oii\ doublet falls shortwards of our spectral coverage with FOCAS.

Figure~\ref{fig:j17spec} shows spectra from the regions discussed. The spectrum of the putative counter-image C is indistinguishable from those of A and B (similar line ratios and $<$\,20\,km\,s$^{-1}$ velocity difference),
supporting the origin of all three in a single source. The spectrum of arc D, however, 
not only exhibits the very different \oiii-to-\hb\ ratio already seen in the images, 
but is also offset to lower velocity, by 60$\pm$10\,km\,s$^{-1}$. 
Clearly D does not originate from the same part of the
source plane as A, B and C; it is presumably a neighbouring galaxy (at separation $\sim$7\,kpc, for the lens model derived below).
In Figure~\ref{fig:j17oiii} (right), we exploit the velocity offset by 
selecting a slightly bluer and narrower wavelength interval, to 
isolate arc D. (The image position shifts, indicating a velocity gradient along this arc.) Given the Einstein radius of the lens derived in the following section, we should expect a counter-image to D, at a radius of $\sim$0.7\,arcsec, but we cannot confidently identify such a feature in the present data.

Even after excluding the contaminating H$\beta$ source, it is difficult to fit our data for \jb\ using pixelised lensing models. Specifically, we find that the optimization always favours slightly improving the fit to arcs A and B, at the expense of not producing any counter-image at C. This is not a failing of the modelling method, but reflects the limitations of our data for this system, e.g. under-sampled PSF and strongly correlated pixels (after rebinning onto square pixels), due to the lack of dithered pointings.

We proceed on the assumption that C is genuinely a counter-image to A and B, 
and crudely enforce this `qualitative' aspect of the model by artificially reducing the noise map values around C, to assign it greater weight.
With this approach, we obtain acceptable fits to a SIE with geometry  (centre, axis ratio and position angle)
 fixed to the lens light, and a sizeable shear of amplitude $\sim$12\,per cent (representing angular structure in the mass not captured in the SIE, and/or external perturbations). 
For a pixelised source, we obtain $\theta_{\rm Ein}$\,=\,1.62$\pm$0.01\,arcsec. Fitting with a parameterized Sersic source  yields  $\theta_{\rm Ein}$\,=\,1.64$\pm$0.01\,arcsec. We adopt the average of these values 
in Section~\ref{sec:mlcalc}. However, to reflect the greater systematic uncertainty in modelling \jb, we impose a 10 per cent error on   $\theta_{\rm Ein}$; deeper and (especially) cleaner data are still required for this lens.

\section{Mass-to-light ratios}\label{sec:mlcalc}

Here we 
derive estimates of the total mass-to-light ratio within the Einstein radius, $(M/L)_{\rm Ein}$, for each lens, and (with assumptions for the dark-matter contribution, and the galaxy age) the stellar mass-to-light ratio, $\Upsilon$, and the IMF mass-excess factor, $\alpha$.

The first part of Table~\ref{tab:calcs} details the observed parameters used in the calculations, i.e. the lens and source redshifts, the Einstein radius and the Einstein-aperture magnitudes in the $i$ band, as measured from the SDSS imaging. For reference, we also note the effective radius and velocity dispersion. We have applied corrections of $\sim$0.1\,mag for the $\sim$1\,arcsec SDSS PSF,
derived from fits with {\sc galfit} \citep{2010AJ....139.2097P}.  
The second part of the table reports the projected Einstein radius in kpc, and the  corresponding projected mass ($M_{\rm Ein}$) and luminosity ($L_{\rm Ein}$). The luminosities assume galactic extinction corrections of $\sim$0.04\,mag in $i$  from \cite{2011ApJ...737..103S}, a band-shifting correction of $\sim$0.09\,mag derived using {\sc EzGal} \citep{2012PASP..124..606M}, and a solar absolute magnitude of $M_{\odot,i}$\,=\,4.53 from the same source. 

\begin{figure*}
\begin{center}
\vskip -1.5mm
\includegraphics[width=165mm]{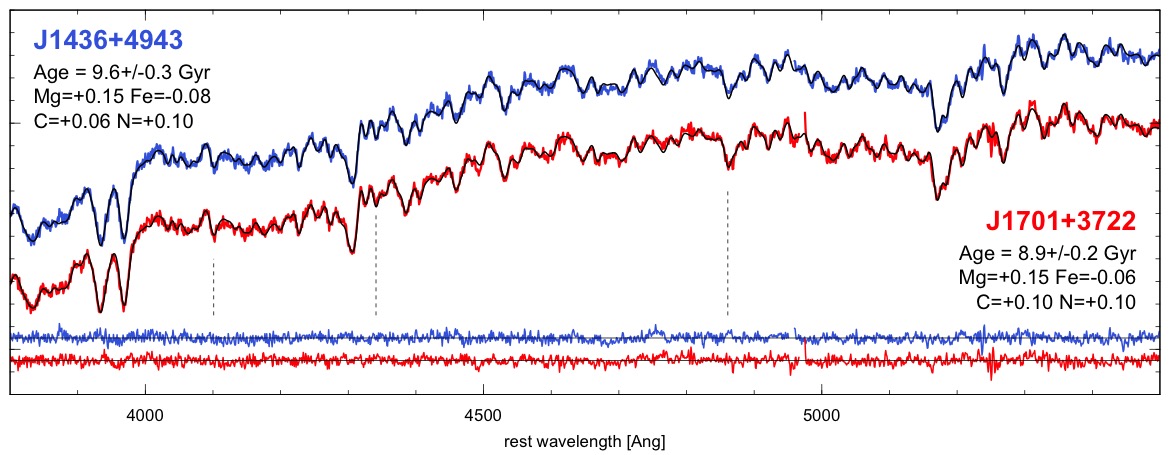}
\end{center}
\vskip -2mm
\caption{Einstein-aperture spectra for the lens galaxies extracted from the MaNGA datacubes, showing the blue spectral range which is most sensitive to the stellar population age. The fits shown in black were derived from full-spectral fitting to the Conroy \& van Dokkum (2012) simple stellar population models. The legend indicates the fitted ages, and the element abundances [Mg/H], [Fe/H], [C/H] and [N/H]. The fit residuals are shown beneath. Note that the spectral fits support old ages for both galaxies. Vertical lines highlight the  age-sensitive H$\delta$, H$\gamma$ and H$\beta$ lines.}
\label{fig:mangaspec}
\end{figure*}

In the third part of Table~\ref{tab:calcs}, we compute the total lensing mass-to-light ratio in the Einstein aperture, which includes contributions from dark matter (and in principle from a central black hole, gas, etc but these are expected to be negligible). 
To compare this with the value expected from the
stars alone, $\Upsilon_{\rm ref}$, we used {\sc EzGal} to derive the latter  for  stellar population models from \cite*{2009ApJ...699..486C}, for metallicities 1.0--1.5$\times$ the solar value and ages 8--12\,Gyr (corresponding to formation redshift $z_{\rm f}$\,$>$\,1.5). 
The full range in $\Upsilon_{\rm ref}$ spanned by these models is 2.41--3.46; our calculations adopt this as the 2$\sigma$ interval, i.e. we assume  $\Upsilon_{\rm ref}$\,=\,2.93$\pm$0.26.
These  assumptions are supported by fits of single-burst models from \cite{2012ApJ...747...69C} to the Einstein-aperture spectra from MaNGA,
which favour ages 9--10\,Gyr 
for both galaxies (Figure~\ref{fig:mangaspec}).
If all of the lensing mass is attributed to stars, the derived {\it maximum} IMF mass factor  $\alpha_{\rm max}$\,=\,$M/(L\Upsilon_{\rm ref})$, is intermediate between the 
values of $\alpha$\,=\,1.0 for a Kroupa IMF and  $\alpha$\,$\approx$\,1.65 for an unbroken power-law IMF with the \cite{1955ApJ...121..161S} slope.

The final part of Table~\ref{tab:calcs} records the correction for dark matter, and its effect on the derived stellar mass-to-light ratio, $\Upsilon$. The corrections are based on the scheme described in \cite{2015MNRAS.449.3441S}, which uses  the projected dark-matter profiles in the EAGLE cosmological simulation, for halos hosting galaxies with $\sigma$\,$\gtrsim$\,250\,km\,s$^{-1}$ \citep{2015MNRAS.451.1247S}.
The estimated contributions are 35$\pm$6 per cent of 
$M_{\rm Ein}$ for the MaNGA systems, roughly twice the fraction obtained for the much closer Smith et al. lenses. (The error here derives from the scatter among halos in the simulation.) After subtracting the halo contribution, we arrive at values for the
stellar mass-to-light ratio $\Upsilon$ and the IMF mass-excess factor
$\alpha$\,=\,$(M-M_{\rm DM})/(L\Upsilon_{\rm ref}$)\,=$\Upsilon/\Upsilon_{\rm ref}$. For \ja, the
resulting $\alpha$ is consistent with unity, i.e. the lensing mass is compatible with a Kroupa IMF and the dark matter expected from EAGLE. 
For \jb, the derived result is lighter than the Kroupa IMF prediction. Heavier-than-Salpeter IMFs are excluded at the $>$4\,$\sigma$ level.

Our estimates are subject to the usual systematic uncertainties affecting
this method. For example, using \cite{2005MNRAS.362..799M} models for $\Upsilon_{\rm ref}$ would increase $\alpha$ by $\sim$12\,per cent; adopting the \cite{2014ApJ...794..135B} cosmological parameters would increase $\alpha$ by 3.5\,per cent. The recovered $\alpha$ anti-correlates with the adopted stellar population age; however, consistency with 
a Salpeter IMF ($\alpha$\,$\gtrsim$\,1.6) would require ages 4--5\,Gyr, which 
are excluded by the weakness of the H$\gamma$ and H$\delta$ 
absorption lines.

Although the IMF constraints for the two galaxies studied here are similar to 
those obtained from lower-redshift lenses \citep{2017ApJ...845..157N,2018MNRAS.478.1595C}, we stress that the current results apply at larger radius (3.7--4.7\,kpc, compared to $\sim$2\,kpc), and are  more dependent on the accuracy of the dark-matter correction.

\begin{table}\label{tab:calcs}
\caption{Adopted parameters and derived quantities from the mass-to-light ratio calculations. The aperture magnitude is quoted before and after correction for the PSF. Luminosities and mass-to-light ratios all refer to the rest-frame SDSS $i$ band and include correction for band-shifting and galactic extinction.  $\Upsilon_{\rm ref}$ is the expected stellar mass-to-light ratio for a Kroupa IMF, given our assumptions for age and metallicity. 
}
\begin{center}
\begin{tabular}{lccc}
\hline
quantity & unit & \ja & \jb   \\
\hline   
$z_{\rm lens}$ & & 0.125 & 0.122 \\
$z_{\rm src}$ & & 1.231 & 0.791 \\
$\theta_{\rm Ein}$ & arcsec & 2.03$\pm$0.04 & 1.63$\pm$0.16 \\
$i_{\rm Ein}^{\rm raw}$ & ABmag & 16.62$\pm$0.02 & 16.76$\pm$0.07 \\
$i_{\rm Ein}^{\rm corr}$ & ABmag & 16.51$\pm$0.02 & 16.66$\pm$0.07 \\
$\sigma_{\rm SDSS}$ & km\,s$^{-1}$ & 282$\pm$10 & 258$\pm$8 \\
$\theta_{\rm Eff}$ & arcsec &  3.78 & 4.30  \\
\hline
$R_{\rm Ein}$ & kpc & 4.72$\pm$0.09 & 3.71$\pm$0.37  \\
$M_{\rm Ein}$ & $10^{10} M_\odot$ & 28.2$\pm$1.1 & 18.8$\pm$3.7  \\
$L_{\rm Ein}$ & $10^{10} L_\odot$ & 6.72$\pm$0.10 & 5.56$\pm$0.35 \\
\hline
$(M/L)_{\rm Ein}$ & $(M/L)_\odot$ & 4.19$\pm$0.10 & 3.38$\pm$0.46 \\
$\Upsilon_{\rm ref}$ & $(M/L)_\odot$ & 2.93$\pm$0.26 & 2.93$\pm$0.26 \\
$\alpha_{\rm max}$ &  & 1.43$\pm$0.14 &  1.16$\pm$0.19 \\
\hline
$M_{\rm DM}$  & $10^{10} M_\odot$ & 9.6$\pm$1.8 & 6.6$\pm$1.6 \\
$\Upsilon$ & $(M/L)_\odot$ & 2.76$\pm$0.27 & 2.20$\pm$0.41 \\
$\alpha$ &  & 0.94$\pm$0.13 & 0.75$\pm$0.16 \\
\hline   
\end{tabular}
\end{center}
\end{table}

\section{Summary}\label{sec:concs}

We have presented Subaru observations of two $z_{\rm l}$\,=\,0.12 strong lenses originally identified from the SDSS-IV MaNGA survey. For each of these systems, the increased signal-to-noise and  resolution greatly improve on the previous data. We have definitively established \ja\ as a multiple-image lens, with a secure Einstein radius. 
For \jb, we have confirmed that two background sources are present, as
suspected from MaNGA, and 
removed the contamination from the second source to derive a new lensing model which is more consistent with the velocity dispersion.

After correction for the dark-matter contributions expected from cosmological simulations, the lensing-derived stellar mass-to-light ratios for the two galaxies are consistent with
(or even lighter than) a \cite{2001MNRAS.322..231K} IMF like that in the Milky Way.
At face value, this $z$\,=\,0.12 application of `pure lensing' (without dynamical information) supports the `lightweight' IMFs derived from 
lenses at $z$\,=\,0.03--0.07 \citep{2017ApJ...845..157N,2018MNRAS.478.1595C}, 
albeit at a larger physical radius, and with a greater reliance on the correctness of our estimated dark-matter contributions.

The observations described here demonstrate the suitability of the FOCAS IFU for galaxy-scale lensing science. 
The MaNGA systems were observed as validation targets during a Subaru search for low-redshift strong-lensing ellipticals; the results of the search programme will be presented in a future paper.

\section*{Acknowledgements} 
RJS and JRL were supported by the Science and Technology Facilities Council through the Durham Astronomy Consolidated Grant 2017--2020 (ST/P000541/1); RJS also acknowledges travel funding from  PATT Linked Grant  2019--2020 (ST/S001557/1).
WPC was supported by a STFC studentship (ST/N50404X/1).
We are grateful to Kentaro Aoki, Takashi Hattori and Marita Morris for supporting our observations during the first science run with the FOCAS IFU. We thank James Nightingale for helpful discussions about lens modelling with {\sc PyAutolens}.

Funding for the Sloan Digital Sky Survey IV has been provided by the Alfred P. Sloan Foundation, the U.S. Department of Energy Office of Science, and the Participating Institutions. SDSS-IV acknowledges
support and resources from the Center for High-Performance Computing at
the University of Utah. The SDSS web site is www.sdss.org.
SDSS-IV is managed by the Astrophysical Research Consortium for the 
Participating Institutions of the SDSS Collaboration including the 
Brazilian Participation Group, the Carnegie Institution for Science, 
Carnegie Mellon University, the Chilean Participation Group, the French Participation Group, Harvard-Smithsonian Center for Astrophysics, 
Instituto de Astrof\'isica de Canarias, The Johns Hopkins University, Kavli Institute for the Physics and Mathematics of the Universe (IPMU) / 
University of Tokyo, the Korean Participation Group, Lawrence Berkeley National Laboratory, 
Leibniz Institut f\"ur Astrophysik Potsdam (AIP),  
Max-Planck-Institut f\"ur Astronomie (MPIA Heidelberg), 
Max-Planck-Institut f\"ur Astrophysik (MPA Garching), 
Max-Planck-Institut f\"ur Extraterrestrische Physik (MPE), 
National Astronomical Observatories of China, New Mexico State University, 
New York University, University of Notre Dame, 
Observat\'ario Nacional / MCTI, The Ohio State University, 
Pennsylvania State University, Shanghai Astronomical Observatory, 
United Kingdom Participation Group,
Universidad Nacional Aut\'onoma de M\'exico, University of Arizona, 
University of Colorado Boulder, University of Oxford, University of Portsmouth, 
University of Utah, University of Virginia, University of Washington, University of Wisconsin, 
Vanderbilt University, and Yale University.

\bibliographystyle{mnras}
\bibliography{rjs}

\bsp
\label{lastpage}
\end{document}